\documentclass[preprint2]{aastex}

\shorttitle{Plasma Heating}
\shortauthors{Milligan}

\begin{document}

\title{A Hot Microflare Observed With {\it RHESSI} and {\it Hinode}}

\author{Ryan O. Milligan\altaffilmark{1}}

\altaffiltext{1}{Solar Physics Laboratory (Code 671), Heliophysics Science Division, NASA Goddard Space Flight Center, Greenbelt, MD 20771, U.S.A.}

\begin{abstract}
{\it RHESSI} and {\it Hinode} observations of a GOES B-class flare are combined to investigate the origin of 15 MK plasma. The absence of any detectable hard X-ray emission coupled with weak blueshifted emission lines (indicating upward velocities averaging only 14~km~s$^{-1}$) suggests that this was a result of direct heating in the corona, as opposed to nonthermal electron precipitation causing chromospheric evaporation. These findings are in agreement with a recent hydrodynamical simulation of microflare plasmas which found that higher temperatures can be attained when less energy is used to accelerate electrons out of the thermal distribution. In addition, unusual redshifts in the 2 MK Fe XV line (indicating downward velocities of $\sim$14~km~s$^{-1}$) were observed cospatial with one of the flare ribbons during the event. Downflows of such high temperature plasma are not predicted by any common flare model.
\end{abstract}


\section{INTRODUCTION} 
\label{intro} 

The ``standard'' solar flare model \citep[e.g.][]{kopp76} has energy stored in coronal magnetic fields liberated via the process of magnetic reconnection. This energy is used in both heating the local plasma and accelerating particles. The accelerated electrons stream along reconnected field lines toward the dense chromosphere where HXR emission is produced as they decelerate. This also results in the  heating of the lower atmosphere through Coulomb collisions with ambient electrons. Due to the tenuous coronal plasma above, the resulting pressure gradient drives the heated material up into the loop at velocities that can reach several hundred km~s$^{-1}$. In more energetic events, a localized pressure enhancement in the chromosphere also creates downflows of cooler, denser material below. This process of chromospheric evaporation is revealed observationally by the detection of blueshifted EUV and SXR emission lines. Many studies have been carried out in recent years that clearly support the evaporation model (e.g. \citealt{acto82,mill06a,mill06b,bros07}).

Although generally accepted, there are still some inconsistencies with the chromospheric evaporation model in the literature. In some spatially integrated, high-cadence emission line spectra, (such as the \ion{Ca}{19} line observed using {\it Yohkoh}/BCS), the stationary component of the line is detected at the flare onset and dominates over any subsequent blueshifted emission (\citealt{dosc05}). \cite{feld90} has also highlighted that in XUV images from {\it Skylab} emission is often observed at the tops of the loops early in the flare. These observations suggest that additional heating is taking place in the corona during the initial stages of a flares' energy release, often before the HXR signature of accelerated electrons is detected.

\begin{figure}[!ht]
\begin{center}
\includegraphics[height=8cm, angle=90]{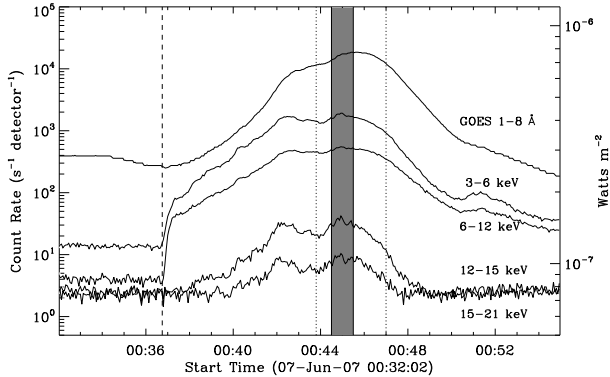}
\caption{{\it RHESSI} lightcurves of the flare in 4 energy bands. Overlaid is the GOES 1--8~\AA~lightcurve. The vertical {\it dashed} line shows when {\it RHESSI} came out of eclipse and the shaded area represents the time over which the {\it RHESSI} images and spectra were taken. The vertical {\it dotted} lines represent the start and end times of the EIS raster taken over the peak of the event.}
\label{hsi_goes_ltc}
\end{center}
\end{figure}

This Letter presents EUV and X-ray observations from {\it Hinode} \citep{kosu07} and the {\it Reuven Ramaty High Energy Solar Spectroscopic Imager} ({\it RHESSI}; \citealt{lin02}) of a GOES B-class flare that appeared to deviate from the standard solar flare model in terms of the origin of the high-temperature plasma. A description of the flare is given in \S~\ref{obs}. The techniques used to analyze the data are described in \S~\ref{data_anal} while the results and their interpretation are given in \S~\ref{results}. A summary is given in \S~\ref{conc}.

\begin{figure*}[!ht]
\begin{center}
\includegraphics[height=16.5cm,angle=90]{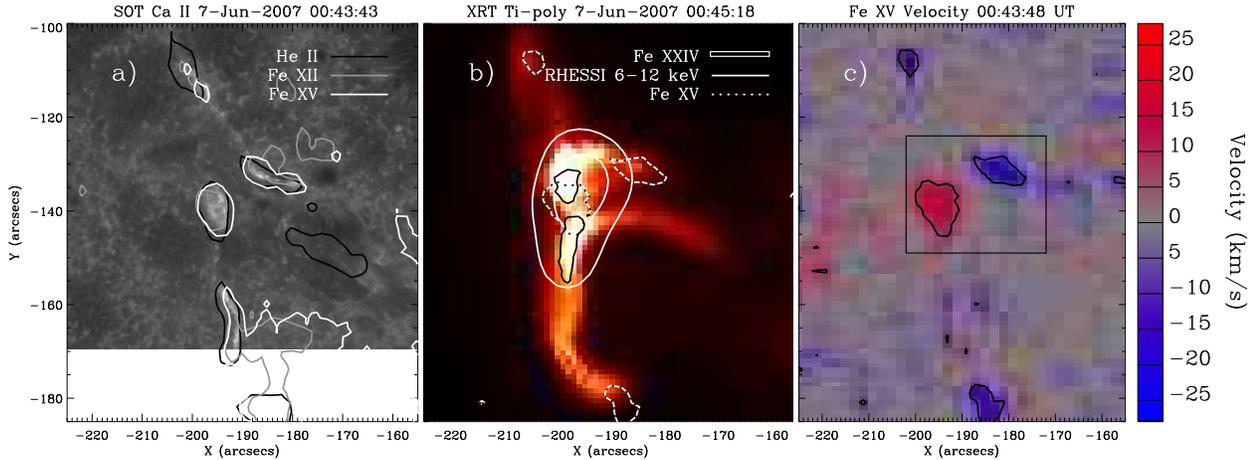}
\caption{({\it a}) An SOT image of the chromosphere taken in the \ion{Ca}{2} H line near the peak of the event. Overplotted are the contours of the \ion{He}{2} ({\it black}), \ion{Fe}{12} ({\it grey}), and \ion{Fe}{15} ({\it white}) emission lines from EIS. The contours are at 45\% of the peak intensity in each image. ({\it b}) An XRT Ti-poly image taken at the X-ray peak of the flare. Overplotted in {\it solid black} lines are the 67\% contours of the \ion{Fe}{24} emission observed by EIS. The {\it solid white} contours are at 50\% and 80\% of the peak of the 6--12~keV source observed by {\it RHESSI} using detectors 3, 4, 6, and 8. The {\it black} and {\it white dotted} lines represent the \ion{Fe}{15} emission. ({\it c}) Velocity map taken in the \ion{Fe}{15} line using EIS. Blue and red pixels represent plasma moving towards and away from the observer, respectively. The box in the center of the image denotes the location of the oppositely directioned flows; the {\it solid black} contours denote the +9 and $-$9~km~s$^{-1}$ levels.}
\label{flare_fig}
\end{center}
\end{figure*}

\section{OBSERVATIONS}
\label{obs}

The observations presented here are of a {\it GOES} B7.6 class flare which began at 00:37~UT on 2007 June 7. The flare originated in the core of NOAA AR10960, a $\beta\gamma$ region close to disc center (49", -117"). The {\it RHESSI} and {\it GOES} lightcurves for the event are shown in Figure~\ref{hsi_goes_ltc}. In the 3--6 and 6--12~keV energy ranges the lightcurves show slowly varying profiles suggesting a predominantly thermal source, whereas the 12--15 and 15--21~keV lightcurves are more impulsive, suggestive of emission from nonthermal electrons. The EUV Imaging Spectrometer (EIS, \citealt{culh07}) observing study used in this work (called {\sc chromo\_evap\_rm}) was designed to search for velocity signatures of chromospheric evaporation at high temperatures (0.05--15~MK) with a relatively high rastering cadence. The Solar Optical Telescope (SOT, \citealt{tsun08}) observed chromospheric emission in the \ion{Ca}{2} H line while X-Ray Telescope (XRT, \citealt{golu06}) provided high-resolution X-ray images.

\begin{figure}[!hb]
\begin{center}
\includegraphics[height=8cm, angle=90]{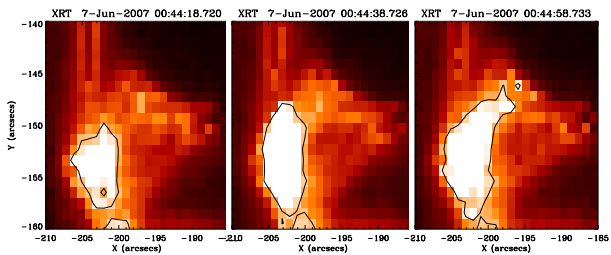}
\caption{A sequence of XRT images taken over 40s around the peak of the flare. A close-up of the northern X-ray loop in Figure~\ref{flare_fig}b shows the apparent flow of hot material along the loop.}
\label{xrt_vel_fig}
\end{center}
\end{figure}

Figure~\ref{flare_fig} shows the emission imaged by {\it RHESSI} and the three {\it Hinode} instruments over the peak of the flare. Figure~\ref{flare_fig}a shows a pair of ribbons observed by SOT in the \ion{Ca}{2} H line. Each ribbon was shown to be associated with regions of opposite polarity on either side of a magnetic neutral line from magnetogram data. Predictably, the \ion{He}{2} (0.05~MK) emission from EIS also mapped out these ribbons. However, emission from hotter lines in the EIS study (\ion{Fe}{12} at 1.25~MK and \ion{Fe}{15} at 2~MK, in particular) still appeared to be aligned with the \ion{He}{2} emission. In non-flaring conditions, this 1--2~MK emission would be in the form of coronal loops rather than from the footpoints. During this event, however, loop emission is only observed above 5~MK as shown in the XRT, EIS (in the 15~MK \ion{Fe}{24} line) and {\it RHESSI} images of Figure~\ref{flare_fig}b. EIS velocity maps in the \ion{Fe}{15} line were constructed to show the spatial distribution of up- and downflowing material. Figure~\ref{flare_fig}c shows the velocity map from the 3-min raster during the X-ray peak of the event. Of particular interest are the two regions of $>$9~km~s$^{-1}$ emission, one redshifted and the other blueshifted, located in the central box. These regions are cospatial with the ribbons in the center of the SOT image and with what appear to be the footpoints of a hot loop observed by XRT. Figure~\ref{xrt_vel_fig} shows a sequence of XRT images that appears to show a build up of material at the eastern leg of a loop and/or the motion of that material over the loop towards the western footpoint. This motion would be in the opposite direction to that expected from the blue- and redshifted \ion{Fe}{15} line of the lower temperature plasma seen apparently at the footpoints of this same loop.

\section{DATA ANALYSIS}
\label{data_anal}

\begin{figure}[!ht]
\begin{center}
\includegraphics[height=8cm, angle=90]{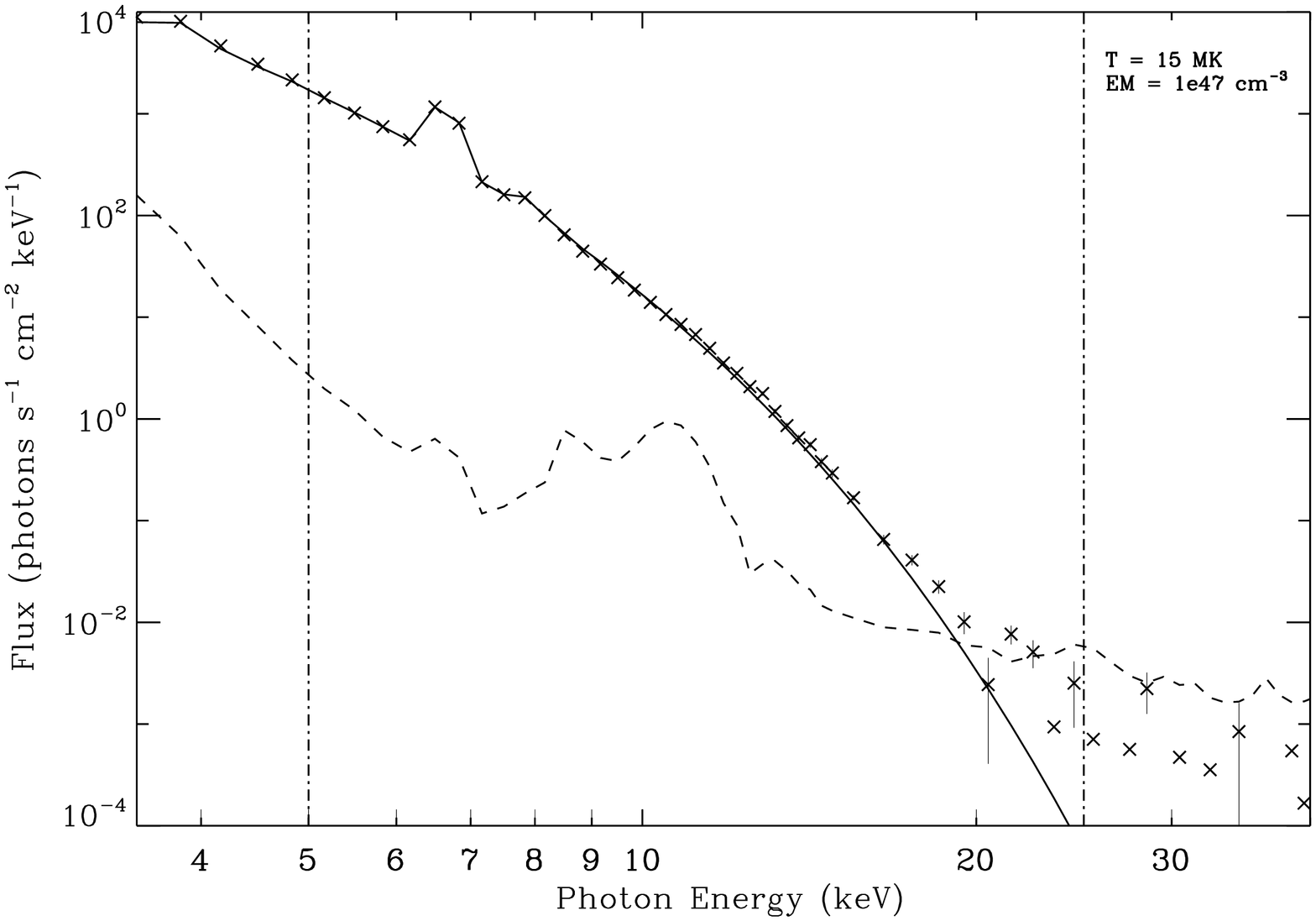}
\caption{Plot of the {\it RHESSI} photon spectrum for 60 seconds at the peak of the flare from 00:44:30--00:45:30~UT obtained from the detector 1 count-rate spectrum. Data points are shown as crosses with $\pm$1$\sigma$ error bars on the flux. The {\it solid} line represents the thermal continuum used to fit the data after correcting for pulse pile-up, while the {\it dashed} line represents the background spectrum. The {\it dot-dashed} lines indicate the 5--25~keV energy range over which the spectrum was fitted.}
\label{hsi_spec_fig}
\end{center}
\end{figure}

At the time of these observations, {\it RHESSI} had been in orbit for over five years and radiation damage had degraded the performance of each of the nine germanium detectors. The effective sensitive area of each detector segment was reduced by poorly known amounts, and the energy calibration was uncertain, particularly at X-ray energies below 25~keV that were critical for this study. Detector \#1 was chosen as the one with the least damage since its count-rate spectrum for the flare of interest here matched similar spectra of flares observed earlier in the mission in the A0 attenuator state \citep{smit02}. The effective detector area is important for determining the absolute incident photon flux and for estimating the emission measure of a thermal spectrum. But the temperature estimates are of greater importance for this study and they depend critically on the detector energy calibration. Reliable corrections to the nominal energy calibration for detector \#1 were made by fitting the iron-line complex seen in the specta of this and many similar flares \citep{denn05}. Since the mean energy of this complex of lines is accurately known from the CHIANTI atomic database to be at 6.7~keV \citep{phil04}, this procedure allows the energy calibration to be determined at that energy to better than $\sim$0.1~keV, corresponding to a temperature uncertainty of 1~MK. Calibration at higher energies was achieved by fitting various germanium lines in the background spectrum, the most notable at $\sim$11~keV seen in Figure~\ref{hsi_spec_fig}, and by comparison with observations made earlier in the mission.

The inferred photon spectrum for one minute over the peak of this event is shown in Figure~\ref{hsi_spec_fig}. In terms of the observed photon energy, few counts were detected above the pre-flare background at energies greater than 20~keV. After correcting for the gain offset and pulse pile-up \citep{smit02} in the count spectrum it was well fitted with the line-plus-continuum spectrum predicted by CHIANTI for an isothermal plasma with a temperature of 15$\pm$1~MK and an emission measure of 2$\times$10$^{47}$~cm$^{-3}$. The iron abundance required to give the best fit to the count-rate spectrum was 1.5 times the photospheric value but the uncertainty on this number is large (possibly as large as a factor of two) because of contamination from the tungsten L-shell lines and other instrumental effects.

EIS has extremely high Doppler velocity resolution ($\sim$3~km~s$^{-1}$), but absolute velocity measurements require that several effects must be taken into consideration. The instrument has an orbital variation in wavelength due to temperature fluctuations. However, in this study, this variation was considered negligible as the raster duration (3~minutes) was short compared to the orbital period ($\sim$3\%). The EIS slit is not oriented perpendicular to the CCD resulting in an additional wavelength variation along the slit. This tilt was accounted for in the analysis, as well as the pointing offset between the short- and long-wavelength detectors. As the rest wavelengths of many lines are not well calibrated, independent measurements needed to be made. In the case of the \ion{Fe}{15} line, an average velocity from a 20$\arcsec \times$20$\arcsec$ region of quiet Sun was used as the rest wavelength. The line profiles appeared symmetrically Gaussian and as such, the velocity for each pixel in the raster was calculated by fitting each line with a single Gaussian profile and measuring the Doppler shift relative to this quiet Sun value. Doppler measurements of the \ion{Fe}{24} line could not be made due to the difficulties in obtaining a reliable rest wavelength.

\section{RESULTS AND DISCUSSION}
\label{results}

\subsection{Plasma Heating}
\label{plasma_heating}

The temperatures derived from the {\it RHESSI} spectral data are significantly high compared to the average values for a B7.6 flare. \cite{feld96} showed that the average {\it GOES} temperature for such a flare would be $\sim$10~MK. The recent statistical analysis of 25\,000 {\it RHESSI} microflares carried out by \cite{hann08a} found that they had temperatures varying from 10~MK, the lowest to which {\it RHESSI} is sensitive, to as high as $\sim$20~MK, with a median value of 13~MK. The detection of plasma at 15~MK is not unusual in itself except that the {\it RHESSI} spectra did not exhibit any hard X-ray emission indicative of nonthermal electrons (see Figure~\ref{hsi_spec_fig}). This, coupled with the relatively low velocity of 14~km~s$^{-1}$ derived from the blueshifted \ion{Fe}{15} emission line, suggests that electron beam driven chromospheric evaporation was not significant during this event. A more likely scenario is that the majority of the energy released during the reconnection process was used to directly heat the local coronal plasma to extremely high temperatures, albeit with a low emission measure. This would have generated an intense downward heat flux resulting in evaporation due to thermal conduction, raising the coronal emission measure to the levels detected by {\it RHESSI}. Had a larger fraction of the energy been used to accelerate particles, the initial coronal temperature may have been considerably lower, reducing the temperature of the conductively evaporated plasma. In the case of beam driven evaporation, the material evaporated by nonthermal electrons would have been much denser than the directly heated coronal plasma and therefore could also not have achieved as high a temperature.

This idea has recently been implemented in a 0D hydrodynamic model of microflare plasmas by \cite{klim08} using Enthalpy-Based Thermal Evolution of Loops (EBTEL). This model determines the temperature, density, and pressure profiles for a flare loop in response to both thermal and nonthermal heating mechanisms. In a sample case, it was found that for a purely thermal heating mechanism, the DEM curve exhibited emission above 3~MK that was absent when half the injected energy was used to accelerate electrons to a mean energy of 50~keV. The authors attribute this to the fact that, while the amount of evaporated material depends on the total energy that is released, the peak temperature is more dependent on the form of the energy. This implies that whenever less energy is used to accelerate electrons, a higher continuum temperature can be achieved by directly heating the local plasma at the reconnection site. 

\subsection{Chromospheric Evaporation.}
\label{chromo_evap}

The histogram of the line-of-sight Doppler velocities for the quiet Sun part of the central box of Figure~\ref{flare_fig}c shows that values followed a Gaussian distribution centered at 0~km~s$^{-1}$ with 3$\sigma$~levels at $\pm$9~km~s$^{-1}$. The mean velocities of the red- and blueshifted \ion{Fe}{15} sources within the box were found to be +14 and $-$14~km~s$^{-1}$, respectively. Each had a $\pm$4~km~s$^{-1}$ 1$\sigma$ dispersion. Throughout the entire event the largest blueshift detected for a single pixel was $\sim$45~km~s$^{-1}$. These relatively low upflow values are often associated with a gentle evaporation process, due either to a low flux of nonthermal electrons \citep{mill06b}, or to a steep temperature gradient along the loop \citep{anti78}. However, the detection of redshifts in emission from 2~MK plasma at a loop footpoint presents a challenge to the chromospheric evaporation model. 

The presence of redshifted material at a footpoint is a signature of explosive evaporation as plasma is driven downwards by the pressure of evaporating material above it in order to satisfy momentum balance  (e.g. \citealt{canf87}). However, both models \citep{fish85} and observations \citep{mill06a,delz06} show that these downflows only occur at temperatures $\lesssim$1~MK in response to a large flux of nonthermal electrons. The absence of any HXR emission here implies that any evaporation observed must be due to a thermal heat flux. It is unlikely that a thermal conduction front could be responsible for these redshifts as its energy would be distributed throughout the transition region and would not cause the required localized pressure enhancement. If the combination of the red- and blueshifted \ion{Fe}{15} emission is interpreted as an end-to-end flow from right to left along the loop, this can be driven by a heat flux  if the energy release is both gradual and asymmetric. These conditions are required to build up sufficient pressure at one footpoint \citep{pats06}. However this interpretation of the flows observed in \ion{Fe}{15} is in contradiction to the perceived flows observed by XRT. Figure~\ref{xrt_vel_fig} shows a sequence of three XRT images during the time of the EIS raster and appeared to show a flow of material from east to west (left to right), although this may actually have been a build up of hot material being deposited in the eastern loop leg which emanated from the western footpoint as the flows in \ion{Fe}{15} actually suggest. 

\section{CONCLUSIONS}
\label{conc}

Observations of a hot microflare are presented using data from {\it RHESSI} and {\it Hinode}. An isothermal fit to the X-ray spectrum at the peak of the flare implies a temperature of 15$\pm$1~MK; a high value for such a relatively weak event. The absence of any significant nonthermal component to the X-ray spectrum measured with {\it RHESSI} implies that this high temperature was achieved through localized heating of the coronal plasma at the reconnection site as opposed to particle acceleration and subsequent chromospheric evaporation. This is suggested by the early heating that is indicated by the gradually rising SXR flux seen by {\it GOES} and {\it RHESSI} starting some 7 minutes before the more impulsive peaks when the blue- and redshifts were observed. Although low-velocity evaporation was observed (mean value of 14~km~s$^{-1}$ in \ion{Fe}{15}), this was likely a consequence of the thermal conduction flux from the plasma heated in the corona. This conduction-driven evaporation was necessary to supply material to the corona as it is unlikely that the preflare density was high enough to result in the observed emission measure at this high temperature.

These findings are consistent with the recent predictions made by \cite{klim08} using their EBTEL model. They investigated how microflare plasma parameters depend on how the injected energy is divided between direct heating and particle acceleration. It was found that by devoting less energy to electron acceleration during micro- and nanoflare events, directly heated plasma reached higher temperatures (see Example 5 in their paper). \cite{hann08b} performed a similar analysis of a {\it GOES} A-class event also using {\it RHESSI} and {\it Hinode} observations. Temperatures comparable to those presented here were determined, but in their case, HXR emission was detected up to 50~keV with a relatively hard spectrum suggesting a significant nonthermal component. The observations presented here show that high-temperature flare plasma can be produced without the need for significant nonthermal electron heating. This result could have a significant influence on flare heating models and the interpretation of future flare observations.

\begin{acknowledgements}
This research was supported by an appointment to the NASA Postdoctoral Program at Goddard Space Flight Center administered by Oak Ridge Associated Universities through a contract with NASA. {\it Hinode} is a Japanese mission developed and launched by ISAS/JAXA, collaborating with NAOJ as a domestic partner and NASA and STFC (UK) as international partners. ROM thanks a number of people for their very helpful and insightful discussions, in particular B. R. Dennis, J. A. Klimchuk, I. G. Hannah, C. L. Raftery, L. K. Harra, D. R. Williams and the {\it RHESSI} and {\it Hinode}/EIS teams.
\end{acknowledgements}

\end{document}